# The meaning of social complexity: insights from a theoretical treatment of the social brain hypothesis

Etienne Lein [a, b], Cécile Aprili [c], Jayaditya Deep [d], Isaac Kazuo Uyehara [a, e]

[a] Max Planck Institute of Animal Behavior, Konstanz, Germany.

[b] University of Konstanz, Germany.

[c] Ecole Nationale Supérieure des Mines de Nancy, France.

[d] Princeton University, New Jersey, USA.

[e] Centre for the Advanced Study of Collective Behaviour, University of Konstanz, Germany.

Corresponding author:

Isaac Kazuo Uyehara

(+1) 530-231-2312

ikuyehara@ucdavis.edu

Department of Viticulture and Enology

UC Davis

595 Hilgard Ln., Davis, CA 95616, USA

Word count of main text: 9370

Since its popularization in the late 1990s (Barton & Dunbar 1997; Dunbar 1998), the Social Brain Hypothesis (SBH) has become a widely influential, but much debated, concept in the field of comparative cognition. Originally, the SBH was proposed to explain why primates possess much larger brains than what would be expected based on their body size, with the prime driver being the apparent complexity of the social environment of this taxonomic order. Species with more socially complex environments were thought to have developed cognitive mechanisms to facilitate navigating a challenging social environment (Byrne & Whiten 1988; with 'group size' as a tentative proxy), resulting in a physical manifestation in the apparatus of central importance to cognition – the brain.

Although the SBH was successful in demonstrating a positive correlation between group size and the relative size of the neocortex in primates (Dunbar 1994; Dunbar 1998), attempts were soon made to extend the SBH to other taxonomic groups – first to non-primate mammalian taxa such as ungulates (Pérez-Barbería & Gordon 2005; Shultz & Dunbar 2006) and carnivores (Pérez-Barberia et al., 2007, Sakai et al., 2011); and later to birds (Emery et al., 2007, West 2014), fishes (Pollen et al., 2007), cetaceans (Fox et al., 2017), and reptiles (De Meester et al., 2019). Along with its application to more species, the SBH was also modified to include additional measures of social complexity outside of group size, including the formation of aggregations, behavioral innovation, social learning, tool use, transactional relationships, alliance formation, caregiving, and cooperation (Reader and Laland, 2002; Burish et al., 2004; Pérez-Barbería & Gordon 2005, Fox et al., 2017). However, these studies only partially reproduced the positive relationships initially found in primates, and some studies reported inconclusive or even opposite results (e.g., Kverková et al., 2018). In parallel, or as a direct result of this – *in toto* – ambiguous body of work, alternative ecological explanations have come more into focus to explain variation in brain size (van der Bijl & Kolm 2016; DeCasien et al. 2017; for a recent review see Heldstab et al. 2022).

The ongoing debate surrounding the SBH is partly due to the inherent difficulty of using correlations between the brain sizes of extant species and measures of their social complexity to infer the existence of past selective pressures. Since researchers are unable to analyze the results of a controlled experiment in brain evolution, it is difficult to identify which variables were responsible for driving brain evolution and which are the constraints or outcomes of brain evolution, especially since these socioecological variables are often correlated. Furthermore, it is not clear which brain metrics, if any, are a suitable indicator for cognitive abilities (Logan et al., 2018). This has led several researchers to believe that comparative analyses of the SBH are highly sensitive to the details of the dataset and analysis, contributing to the lack of robust findings (Powell et al., 2017; Wartel et al., 2019; Hooper et al., 2022). In response, it has been suggested that these issues can generally be resolved by properly formulating hypotheses and using appropriate statistical techniques (Dunbar and Schultz, 2023).

While the inherent challenges associated with comparative analyses have no doubt contributed to the controversy surrounding the SBH, we believe that the SBH is also prone to particularly inconclusive results due to inconsistent conceptual and operational definitions of social complexity (Kappeler, 2019). Here, we introduce a theoretical model as a thought experiment to clarify the conceptual definition of social complexity when testing the SBH and demonstrate the difficulties and potential solutions to quantifying social complexity. To overcome the challenges associated with quantifying social complexity, we believe that SBH research should adopt a hypothesis-appropriate conceptual definition of social complexity, take a more cautious approach to quantifying social complexity, and expand research into the study of cognition in social situations.

**THOUGHT EXPERIMENT**

We present a thought experiment to help reveal a conceptual definition of social complexity and illustrate how, even in this simplified system, quantifying social complexity from first principles is extremely difficult. In this thought experiment, we imagine a group of species that differ only in their social environments and ask which species should have the highest levels of cognitive investment. As opposed to a more realistic comparative analysis, our thought experiment allows us to assume that systematic differences in cognitive investment across species must be attributable to their social environment, as this is the only way in which the species differ. To avoid issues related to measurement error, we also assume that we can perfectly measure the social environment of different species and perfectly measure the metric of cognitive investment relevant to social interactions.

Suppose that all individuals of a species are embedded in a social network and that these social networks can be fully defined by the number of individuals in the network (N), the number of other individuals that a focal individual interacts with in the network ($P$), and the total number of lifetime conspecific interactions a focal individual has ($I$). To further simplify this thought experiment, we assume that all individuals of a species exist in identical social networks and that interactions between a focal individual and all of its interaction partners are equally likely. This corresponds to each species having a social network characterized by $N$ nodes, with each node having $P$ edges, and each edge weighted by $I/P$. For example, a species could be characterized by always living in groups of ten ($N$ = 10), with each individual only interacting with six other group members ($P$ = 6), and each individual interacting with its six interaction partners a total of 180 times during its life ($I$ = 180), yielding on average, 30 lifetime interactions per interaction partner.

To eliminate other complicating factors, we assume that all of these species share a single interaction type and all individuals use the same cognitive process to make decisions during this interaction. We imagine that each species can be represented as social groups that live together and compete for resources following the hawk-dove model (Smith, 1982). In the hawk-dove model, two interacting individuals can play either the hawk strategy or the dove strategy to gain access to a resource. If both individuals play hawk, they engage in a contest for the resource, with the winner getting to use the resource and both individuals paying some costs associated with winning or losing the contest. If one individual plays hawk and the other plays dove, the hawk player gets full access to the resource and the dove player gets no access to the resource. If both individuals play dove, they share the resource equally. The hawk-dove model is one of the simplest game theory treatments of animal contests because the evolutionarily stable strategy (ESS), which represents the optimal strategy to this game, can be calculated as a function of the payouts of the different interaction outcomes. This means that we can solve the relevant equations that allow us to predict the frequency at which an individual will play hawk versus dove following evolution.

To extend the classic hawk-dove model, we assume that individuals have a unique resource holding potential, which represents their fighting ability and dictates the winner of a hawk-hawk interaction (Parker 1974). If two individuals compete for a resource and both choose the hawk strategy, the individual with the higher resource holding potential will win the contest. To adapt the hawk-dove model to study the SBH, we also allow individuals to invest in increased cognitive abilities that allow them to avoid playing hawk when they have a lower resource holding potential than their opponent and to avoid playing dove when they have a higher resource holding potential than their opponent. To represent the decision-making process of an individual with an explicit and simple cognitive mechanism, we imagine that each individual can remember the outcomes of previous hawk-dove games. In this framework, we assume that individuals have no knowledge of their own resource holding potential, but can perceive and remember the resource holding potential of opponents and remember whether they won a hawk-hawk interaction against that opponent. Thus, an individual would never play hawk against an opponent if it remembers losing a hawk-hawk contest to an individual with equal or lower resource holding potential. Similarly, an individual would never play dove against an opponent if it remembers winning a hawk-hawk contest against an opponent of equal or larger resource holding potential. When an individual cannot infer that it would definitely win or lose a hawk-hawk interaction with an opponent, it must randomly decide to play hawk or dove.

We assume that different individuals can also have different memory capacities, which correspond to the number of their previous interactions they can remember at one time. Individuals with larger memory capacities are able to remember more of their previous memories, which should give them more information when making a decision to play hawk or dove, and generally reduce their probability of playing a suboptimal strategy (Fig 1A). Finally, we impose the constraint that individuals with a higher memory capacity must also pay an increased lifetime metabolic cost for having and maintaining this increased cognitive capacity. Taken together, individuals of a species in our thought experiment all exist in the same social environment, fully characterized by the number of individuals in their group ($N$), the number of group members with whom they play hawk-dove games ($P$), and the total number of hawk-dove games they play with their interaction partners ($I$) (Fig 1B-D).

This simplified treatment allows us to ignore issues associated with measuring cognitive investment or the social environment, analyzing cognitive tasks that depend on several interacting cognitive abilities, and quantifying social environments with more dimensions of social complexity, such as heterogeneity among individuals and changes in social environment over time. Now we ask whether, in these idealized conditions, we can *a priori* define social complexity and rank species in order of social complexity if we know their $N$, $P$, and $I$. If we can sufficiently quantify the social complexity of different species in this thought experiment, we could then directly test the SBH by comparing the social complexity and memory capacity of different species. In this case, we would be confident that a strong positive correlation between the social complexity and memory capacity is evidence that social complexity drove selection for increased memory capacity, while no correlation or a negative correlation would be evidence against the SBH. On the other hand, if we cannot sufficiently define and measure social complexity, we would not be able to confidently test the SBH, as we would have no reliable measure of social complexity to regress with memory capacity. We believe that this thought experiment represents an ideal evolutionary scenario for a researcher testing the SBH and we propose that if we cannot define and measure social complexity in this thought experiment, we are unlikely to be able to do so in real systems. In the following, we will use this thought experiment to arrive at a conceptual definition of social complexity and evaluate to what extent we can quantify social complexity.

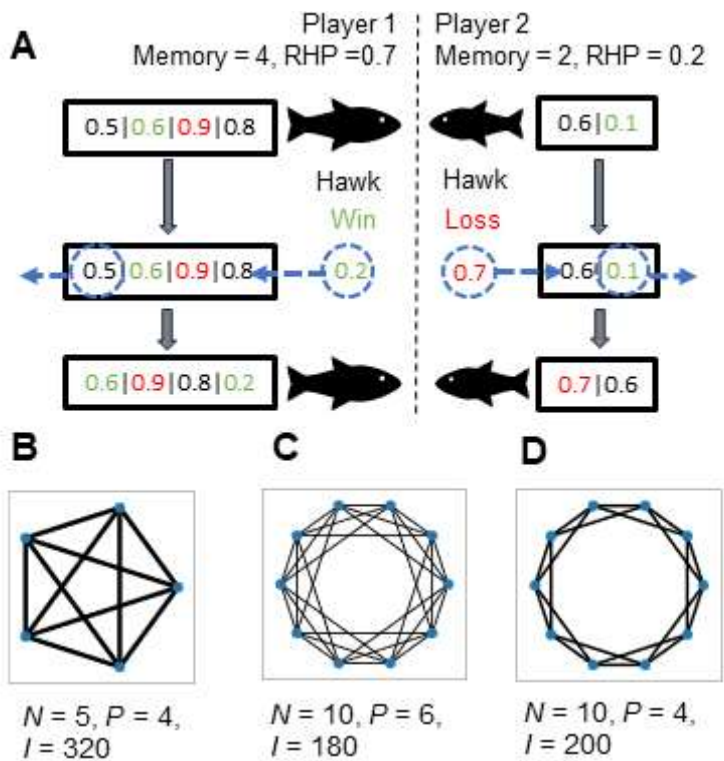

Fig. 1. A) An example of a hawk-dove game played between two players. Memories are represented by the RHP of the opponent played, with green memories indicating hawk-hawk victories, red memories indicating hawk-hawk losses, and black memories indicating hawk-dove or dove-dove interactions. Since Player 2 is smaller than an opponent Player 1 remembers beating in a hawk-hawk contest, Player 1 will play hawk. Player 1 is larger than an opponent Player 2 remembers beating and Player 2 does not remember losing any hawk-hawk contests, so Player 2 will choose to play hawk or dove randomly. In this example, Player 2 chooses to play hawk and loses because it has a smaller RHP than Player 1. This memory is then added to both players memories, causing their oldest memory to be forgotten. B) An example of a social environment with 5 total individuals ($N$), four interaction partners ($P$), and an interaction frequency of 320 ($I$). individuals are represented by nodes (blue circles), interaction partners are those individuals connected by edges (black lines), and edge width is proportional to the number of interactions two connected individuals will have. In this example, all individuals in the network interact with all other individuals and each individual will play 80 hawk-dove games with each of its partners, for a total of 320 lifetime hawk-dove games. C) An example of a social environment with 10 total

individuals (*N*), six interaction partners (*P*), and an interaction frequency of 180. Compared to individuals in B, each individual interacts with more partners but since the number of interactions per partner is lower (30), each individual plays fewer total hawk-dove games. D) An example of a social environment with 10 total individuals (*N*), 4 interaction partners (*P*), and an interaction frequency of 200. In this social environment, individuals can interact with the same number of individuals as in B, but they play 50 games with each individual as opposed to 80, and are in a community with twice the total number of individuals.

## DEFINING SOCIAL COMPLEXITY

While there are many different ways to define social complexity, each with pros and cons, it is important that tests of the SBH use a consistent conceptual definition of social complexity. By first understanding what social complexity would mean in our thought experiment, we may arrive at a definition that can be applied more generally. In our thought experiment, we constrained the social environment to include only the number of individuals in a group, the number of individuals in that group with whom a single individual interacts, and the number of interactions an individual will have in its lifetime. Although at first glance it seems intuitive that as all three of these dimensions increase, the social complexity of the system should increase, the exact quantitative relationship between social complexity and these variables is not obvious. If we define social complexity with respect to the SBH, social complexity must create selective pressure for increased cognitive investment. In this framework, the SBH proposes that this selective pressure did drive increased cognitive investment in the evolutionary past.

In our example, socially complex environments must therefore increase demand for memory. That demand, must in turn be driven by its marginal value in the decision-making process, which is simply the decision of whether to play hawk or dove. Thus, an increase in memory capacity has value to an individual only in so far as it increases the chances that an individual will correctly play hawk or dove in a hawk-dove game. Consequently, the dimensions of the social environment increase social complexity (Fig. 2A) when it is more difficult for individuals to correctly choose to play either hawk or dove in that social environment (Fig. 2B). For an individual with a given memory, an increase in the social complexity of its environment will have a negative impact on its fitness, as it would be less likely to choose the correct strategy during a hawk-dove game, resulting in lower expected resources (Fig 2C). However, individuals that pay a metabolic cost to increase their memory capacity may perform better in hawk-dove games (Fig. 2D), driving selection for increased memory (Fig. 2E). By generalizing from our thought experiment, we propose that the SBH is a set of hypotheses that suggest: 1) a social environment increases in social complexity when it makes a cognitive task more difficult, 2) the increased cognitive difficulty of the task decreases performance on the task, which then incurs a fitness cost, 3) and this drives selection for an increased investment in a cognitive structure that can reduce these fitness costs (Fig. 2).

Therefore, for purposes of studying the SBH, we define social complexity as the function of the social environment that controls the expected fitness consequences of performance on a relevant cognitive task as a function of cognitive investment. That is, all else equal, as social complexity increases, cognitive investment must also increase to maintain the same performance on a cognitive task. We chose this definition because it directly captures the mechanism underlying the SBH (Fig. 2A-C) and would, by definition, represent the influence of the social environment on the optimal cognitive investment (Fig. 2D-E). This definition aligns with previous arguments that suggest that social complexity should be defined with respect to cognition (Bergman and Beehner, 2015). Social complexity could also be defined more abstractly, for instance, as the qualities of a social environment that make it more difficult to manage social information (Emery et al., 2007). However, we believe that by highlighting a similar concept in a more explicitly evolutionary framework, it is more obvious what would constitute evidence for the SBH. If a researcher could measure social complexity following our definition, we would expect that the absence of a correlation between social complexity and cognitive investment (Fig. 2E) would be strong evidence that the fitness gains of cognitive investment did not outweigh the fitness costs of cognitive investment (Fig. 2D). On the other hand, a positive correlation between social complexity and cognitive investment would support the idea that increased cognitive investment evolved as a direct response to increased social complexity in the evolutionary past.

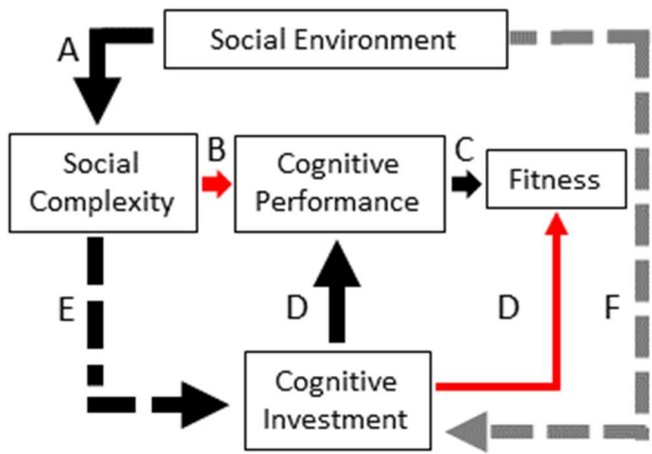

Fig. 2. A conceptual framework showing the general causative relationships hypothesized by the Social Brain Hypothesis, where positive relationships are black, negative relationships are red, and indirect relationships are dashed. The basis for the Social Brain Hypothesis is that the social environment creates social complexity (A) and that this social complexity reduces performance on a cognitive task (B). This reduction in cognitive performance then decreases fitness (C). However, increased cognitive investment can improve performance on the cognitive task, which will result in a net fitness gain when the fitness benefits of increased cognitive investment outweigh the fitness costs of that cognitive investment (D). When an increase in cognitive investment leads to a net increase in fitness, increased social complexity indirectly selects for increased cognitive investment, producing a correlation between social complexity and cognitive investment (E). We should not expect an arbitrary measure of the social environment to be correlated with cognitive investment, as there is no mechanism linking it to selection on cognitive investment (F). Disentangling these relationships in empirical systems is difficult due to potential drivers of cognitive investment outside of social complexity.

Comparative studies of the SBH are essentially looking for empirical evidence of social complexity having exerted an indirect selective pressure on cognitive investment (Fig. 2E). However, a measure of social complexity should be correlated with cognitive investment only in so far as the difficulty of a cognitive task increases with the measure of social complexity (Fig. 2B), the increased difficulty of the cognitive task has negative fitness consequences (Fig. 2C), and increased cognitive investment can improve fitness by decreasing the difficulty of the cognitive task relative to its cost (Fig. 2D). For example, the original SBH explicitly proposed that primates formed groups to decrease predation rate and that the social challenges associated with maintaining group cohesion created selection for an increase in the size of the socially relevant parts of the neocortex (Dunbar, 1998). By creating a framework for understanding how social pressures effected changes in brain evolution, the original SBH also hypothesized why a correlation between group size and neocortex size should be expected – larger groups are more socially complex because the cognitive demand necessary to mitigate the fitness costs of social conflict increases with group size (Dunbar and Schultz, 2023). Thus, by using group size as the measure of social complexity, the original SBH specifically posited that group size determines, or at least constrains, the cognitive investment necessary to avoid a certain level of social conflict in primates. The argument is not that group size *per se* selects for higher cognitive investment, but rather the cognitive challenge posed by group size selects for higher cognitive investment. This example illustrates how all comparative tests of the SBH are not just tests of a general socioecological hypothesis (*e.g.*, primates with increased neocortex size maintain larger group sizes, Fig. 2C-D), but also whether the difficulty of a cognitive task scales with their measure of social complexity (*e.g.*, the challenge of resolving social conflicts increases with group size, Fig. 2A-B).

In our thought experiment, we could imagine that the difficulty of choosing the correct strategy in a hawk-dove game is only a function of the number of interaction partners an individual has. In this case, some function of interaction partners would be our measure of social complexity because only the number of interaction partners affects the difficulty of the hawk-dove game. We could then evaluate whether social complexity was sufficient to drive an increased cognitive investment by determining whether the benefits of increased memory capacity outweighed the metabolic costs of that investment. This would be a straightforward test of the SBH when viewing the thought experiment in an evolutionary framework. For a comparative analysis, we would only have access to the outcome of evolution in the model and we would compare the memory of species in different types of social groups. If the SBH were true in this example, we would expect to see a positive correlation between interaction partners and memory across species. However, if we regressed the total number of individuals in a group against memory, we would not expect there to be a correlation simply because it was interaction partners, not total group size, that affected performance in the hawk-dove game. That is, if our proxies for social complexity are chosen because they are convenient measures of the social environment or they were used in a previous study, the lack of a correlation between social complexity and cognitive investment could be due to the proxy of social complexity having a weak correlation with what we defined here as constituting social complexity. In other words, if the SBH is tested without an explicit evolutionary rationale, there is no *a priori* reason to believe that an arbitrary metric of social complexity should be correlated with an arbitrary metric of cognitive investment in the chosen taxa. Consequently, evidence for or against the SBH is only trustworthy if we are confident that the measure of social complexity tracks the cognitive demand of the social environment.

This is well exemplified in real species when investigating the SBH in social insects. Given the criteria used by other researchers, the eusocial Hymenoptera could be considered more socially complex than solitary Hymenoptera, as they live in larger groups, have division of labor, and breed cooperatively. As it turns out, there is no strong evidence for higher cognitive investment in more social Hymenoptera (Farris & Shulmeister, 2010; O'Donnell et al., 2015), but perhaps this is unsurprising if we consider that the task specialization found in eusocial insects may actually decrease the cognitive demands on individuals (Lihoreau et al., 2012; Pahlke et al., 2021). By framing the SBH more explicitly and using our conceptual definition of social complexity, we are forced to ask what additional demand for cognitive investment, if any, is created by living in a eusocial system, not whether one could argue that eusocial species are more socially complex than solitary species.

**QUANTIFYING SOCIAL COMPLEXITY**

Even if we use a consistent and evolutionarily grounded definition of social complexity, it is a separate task to properly quantify it. In real systems with many socioecological dimensions, we are generally forced to make use of heuristics derived from our expertise with the system. For instance, we could intuit that the relevant metric for group size is not the total number of individuals that are aggregated, but the number of individuals with whom any one individual has bonded relationships. Alternatively, we could categorize certain behaviors as indicative of social complexity and create an index based on the occurrence of those behaviors in a species. These operational definitions are generally constructed based on what we deem to be the biologically relevant components of the social environment and will therefore likely change depending on the judgement of individual researchers and the species of interest. By categorizing behaviors or species based on our own criteria, metrics, and thresholds, we may also be oversimplifying the quantification of social complexity, resulting in a proxy that can no longer reliably rank species based on the cognitive demand imposed by the social environment. This makes it possible for a researcher to inadvertently predetermine the outcome of their test of the SBH based solely on how they decided to construct their measurement of social complexity. In our thought experiment, we have three continuous dimensions of the social environment and, even in this simplified environment, it is not obvious how to rank social networks based on social complexity because network size, interaction partners, and interaction frequency may have counterbalancing effects (Fig. 1B-D). When one network has individuals with twice the interaction partners but half of the interaction frequency as another, which network do we expect to be more socially complex? Are smaller networks where all individuals interact more or less complex than ones with the same number of interaction partners, but each individual only interacts with a small fraction of the total network?

Although the specifics of our thought experiment are not meant to be generalizable, if we cannot intuit how to quantify social complexity in this simple example, we find it unlikely that we will be able to do so in real systems.

*Solving for social complexity*

Since we cannot construct a metric of social complexity for our thought experiment using intuition alone, we attempted to solve for social complexity mathematically. To quantify social complexity in our thought experiment, one would have to quantify the expected payout of a hawk-dove game for an individual as a function of its memory, its social environment, and the memory capacity of the other individuals in its social network. That is, social complexity is some function of our three-dimensional socioecological parameter space that relates how our social network parameters change the memory capacity necessary to maintain a certain success rate when making hawk-dove decisions against other individuals of a given memory. If we account for the metabolic costs of memory, our social complexity function should also help us recover the expected fitness of an individual as a function of its memory and social environment. Having a fitness function would then allow us to potentially calculate the ESS memory for a given social network, similar to how the ESS behavioral strategies can be calculated in the simplified hawk-dove model.

However, in our modified hawk-dove model, the addition of a cognitive component and social component made it too difficult for us to solve for either social complexity or ESS memory based on the payout matrix and social environment variables. We encourage others to try derive an analytic expression for social complexity or the ESS memory in our model or alternative mechanistic presentations of the SBH. Without an expression for social complexity, we were forced to treat it as a black box function that transforms the social environment into the amount of memory it takes to maintain a certain performance in hawk-dove games. In practical terms, that means that if we were analyzing data produced from our simple model, we would be unable to regress social complexity and ESS memory capacity to test the SBH because we have no way to quantify social complexity. Comparative studies implicitly have the same issue anytime they propose that more than one aspect of the social environment can affect social complexity or when variables are categorical rather than continuous.

We find it noteworthy that it was so difficult to mathematically solve for social complexity because we created this minimal presentation of the SBH specifically so that we could solve for social complexity. Even when we knew exactly how individuals made a binary decision, how cognitive investment impacted the decision-making process, and how the outcome of those decisions affected fitness, we could not determine what exactly about the social environment drove demand for cognitive investment. In real systems, we expect that there are more than three ways that the social environment of a species can change, the cognitive mechanism underlying how performance changes as a function of the social environment will be poorly understood, the fitness consequences of performance on some cognitive task will be intractable, and that the costs of cognitive investments will be hard to estimate. Furthermore, different species in a real system are unlikely to have identical ecological parameters, will likely have multiple relevant cognitive tasks that are affected by the social environment, and may use different cognitive and behavioral mechanisms to solve the same tasks. Thus, we believe that finding an analytical expression for social complexity in a given taxon will generally be intractable with our current level of knowledge and, as in our model, social complexity should generally be treated as an unknown quantity.

*Using simple proxies of social complexity*

If we assume that social complexity is likely too complex to properly quantify for testing the SBH, we may be forced to use a univariate proxy of social complexity. Although this approach should be sufficient when there truly is one major variable or axis of social complexity, this may lead to spurious results if we do not correctly identify that social variable or if there are interactions between other dimensions of the social environment and our univariate proxy of social complexity. Under these circumstances, we will likely underestimate the correlation between cognitive investment and social complexity because any error in our estimation of social complexity will weaken the true correlation between the two variables. This is even more important if the strength of correlations is being used to distinguish between social and ecological drivers of brain evolution.

To demonstrate how statistical tests of the SBH are vulnerable to poor inferences when we have no expression for social complexity, we simulated data by modeling the mechanisms in our thought experiment (SI). This involved simulating the evolution of memory capacity for species in different social environments following our model, allowing us to see the outcome of selection in controlled conditions when all dimensions of the social environment are known. Our entire simulated dataset can be considered a representation of a hypothetical taxonomic order that includes 200 species with all possible combinations of our three social variables across their given ranges, while all other aspects of the species can be considered identical and all variables are measured without error – essentially a perfect dataset for studying the SBH. We assume that ESS memories can be used to reliably rank the social complexity across different networks, because individuals with more memory in this model, by definition, should be in more socially complex environments. In our simulations, we found that ESS memory is positively correlated with network size ($N$), interaction partners ($P$), and interaction frequency ($I$) when using the full dataset (Fig. 3A-C). In the following, we show examples of how the relationships between ESS memory and our three social dimensions can all be altered when we no longer have access to the idealized dataset (Fig. 3D-F). By analyzing subsets of the full dataset, we mimic the more realistic scenario in which the studied species span a limited range of values along some social dimensions, the social dimensions of the species are correlated, or some of the relevant dimensions of social complexity are not measured. This highlights how the assumption that social complexity should increase with any of our three social variables can lead to misleading inferences about the SBH because social complexity is, in fact, a complex multivariate function.

First, we show that the significantly positive relationship between network size and ESS memory when using the full dataset disappears if one were only analyzing the 40 species with our lowest interaction frequency ($I$ = 100; Fig. 3D). Although ESS memory generally increases with network size, this relationship is driven by the high correlation between network size and ESS memory in species with a high interaction frequency (*e.g.,* $I$ = 500, $r$ = 0.59, $p$ = 0.001). A researcher studying species with a low interaction frequency (in this example: 100) might infer that the SBH is incorrect because of the lack of a positive relationship between network size and ESS memory. However, with more complete information they might be able to infer that network size does not have a meaningful impact on social complexity when interaction frequency is relatively low, but it has a positive relationship with social complexity when interaction frequency is relatively high. That is, even though this subset of species spans a range of network sizes, it does not vary in social complexity, making it an inappropriate group for studying the SBH. Conversely, another researcher studying a group of species with high interaction frequency would find support for the SBH using network size as their proxy of social complexity. Although both researchers are using the same proxy for social complexity, their difference in conclusions is due to the fact that network size alone is an unsuitable proxy for social complexity. This issue is caused by an interaction between network size and interaction frequency that cannot be captured by univariate proxies of social complexity.

Next, we show that the positive correlation between the number of interaction partners and ESS memory in the full dataset would disappear for a study that happened to only compare the 100 species with at least 16 interaction partners, thus omitting species with four or eight interaction partners (Fig 3E). As a result, a researcher using interaction partners as a proxy for social complexity would not find a correlation between ESS memory and social complexity if their study species had interaction partners ranging between 16 and 64. On the other hand, a researcher using interaction partners as a proxy for social complexity would find support for the SBH if their study included species with a lower number of interaction partners. Hence, the implicit assumption that an increase in a socioecological dimension results in a monotonic increase in cognitive investment would lead to the incorrect conclusion that the number of interaction partners does not affect the evolution of memory *per se*. In this case, this is because there is a non-linear relationship between interaction partners and ESS memory. This is exemplified in cetaceans, where somewhat counterintuitively, brain size is highest in species with intermediate group sizes (Fox et al., 2017).

Finally, we show how the existence of an undetected negative correlation between the number of interaction partners and the interaction frequency could create a negative relationship between interaction frequency and ESS memory (Fig. 3F). If a study happened to include only the 28 species that had either a high number of interaction partners (64) and a low interaction frequency (200-300) or a low number of interaction partners (four) and a high interaction frequency (300-500), this would result in a significant negative correlation between interaction frequency and ESS memory, even though there is a generally positive relationship between interaction frequency and ESS memory in the full dataset. This example illustrates how undetected multicollinearity can lead to spurious results. Avoiding this situation when using real data may be challenging specifically because we are likely unaware of all of the relevant dimensions of social complexity and these dimensions are unlikely to be independent in nature. As in the previous examples, using a single dimension of the social environment as a proxy of social complexity fails to correlate with social complexity due to the complex way that the dimensions of the social environment produce social complexity. For all three of these examples, a more complete quantitative understanding of social complexity would have allowed us to account for the interactions between different dimensions of the social environment and regress social complexity itself against cognitive investment. Without an understanding of the underlying function that maps the social environment to social complexity, it is difficult to know to what extent different comparative analyses are truly regressing social complexity and cognitive investment.

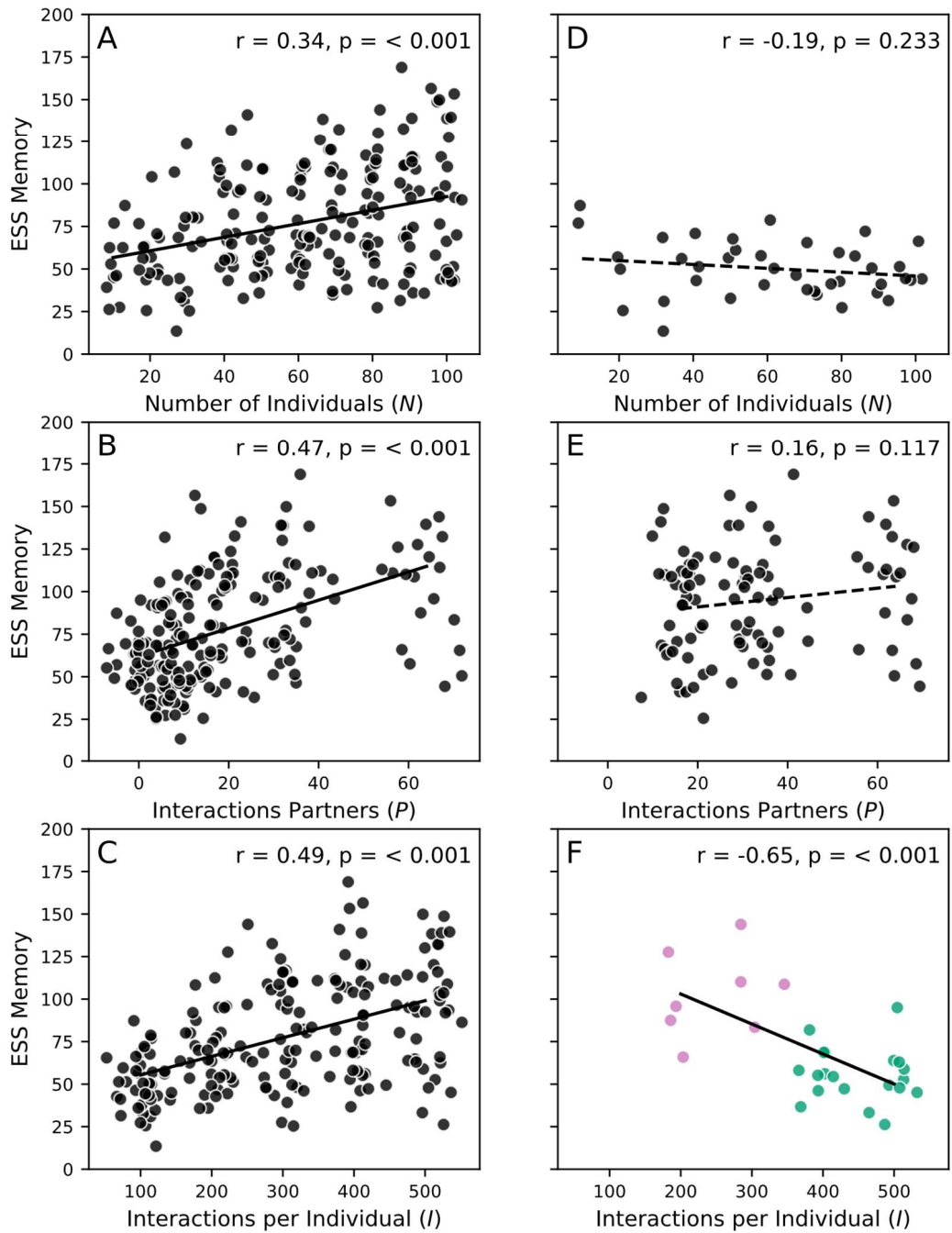

Fig. 3. The ESS memory of all 200 simulated species in the full dataset as a function of their A) number of individuals in the network (*N*), B) interaction partners (*P*), and C) interactions per individual (*I*). While there are positive relationships between each of our social network parameters and ESS memory, these relationships are altered when only a subset of the relevant social dimensions are available or controlled for. D) When we only analyze species with 100 interactions per individual, there is no positive relationship between number of individuals and ESS memory for species. This is because interactions per individual affects the relationship between number of individuals and ESS memory. E) If only species with 16 or more interaction partners are analyzed, there is no longer a significant correlation between interaction partners and ESS memory since the relationship between ESS memory and interaction partners is non-linear. F) If there are species with many interaction partners (64, purple) and few interaction partners (4, green), but the species with many interaction partners have few interactions per individual and species with few interaction partners have many interactions per individual, there is a negative correlation between interaction per individual and ESS memory. Data are jittered along the x-axis to aid visualization.

*Using multivariate proxies for social complexity*

Since univariate proxies of social complexity may be unsuitable when there are multiple dimensions of the social environment that contribute to social complexity, it may be attractive to instead create multivariate proxies of social complexity. In our thought experiment, the true multivariate measure of social complexity would be the analytical expression for social complexity that we were unable to derive. In addition to being unable to derive the exact equation for social complexity, we are also unable to create a *de novo* multivariate equation that can be used to rank social environments based on their social complexity. In a real system, we may be unaware of the relevant social variables or be unable to measure them, yet even with perfect measurements on relevant social variables, it is unlikely that we can correctly combine these variables into a function of social complexity that captures the difficulty of relevant cognitive tasks.

It is particularly difficult to intuit how different aspects of the social environment should contribute and interact to produce socially complexity when those variables must all be converted to a single unit, but are as diverse as group size, social structure (Kappeler, 2019), mating system and degree of parental care (Fischer & Jungwirth, 2022), territoriality (Hardie & Cooney, 2023), and the presence of transactional social interactions (Burish et al., 2004). More nuanced efforts to unify social complexity into a more rigorous *a priori* quantity directly address this by using units that generalize across species and capture a proposed evolutionary link between social complexity and cognition. For example, the number and diversity of differentiated relationships have been proposed as general measures of social complexity (Bergman and Beehner, 2015; Fischer et al., 2017). These definitions focus on a single axis that potentially captures cognitive demand, similar to our conceptual definition of social complexity. When these definitions and metrics are created *a priori*, they also preclude the use of arbitrary post-hoc constructions of social complexity. However, these definitions, and similar unified metrics, must eventually contend with how to properly classify, weigh, and calculate factors such as the number of differentiated relationships, the different degrees of differentiation, and the diversity of relationships in a way that approximates how cognitively challenging it is to maintain those relationships.

For instance, quantifying the diversity of differentiated relationships of an individual has the same complications as calculating species diversity. Since we have multiple widely used ways to quantify species diversity (e.g. species richness, Simpson's Index, and Shannon-Weiner Index) that all differently weigh different aspects of diversity, our inferences about how species diversity affects an ecological community may depend entirely on how we decide to calculate species diversity (Hurlbert, 1971). Similarly, our inferences about how the diversity of differentiated relationships affects brain evolution may also depend on how we mathematically formulate diversity. While there may be an equation for the diversity of differentiated relationships that captures the cognitive demand imposed by social relationships, it is difficult to intuit what that equation should be. Thus, while we believe definitions that are explicitly formulated to represent cognitive demand conceptually align with tests of the SBH and represent the standard for how social complexity should be defined and measured, it is difficult to substantiate that these definitions and metrics can consistently rank species based on cognitive demand.

An alternative approach to creating a multivariate proxy of social complexity *de novo* is to use statistical inference to find the best social predictors of cognitive investment. However, instead of testing the SBH directly, this approach effectively assumes that the SBH is correct and then tries to infer the relevant dimensions of social complexity. Evidence that cognitive investment can be predicted from social variables can therefore only be considered tentative support for the SBH and one must be careful not to affirm the SBH because social complexity predicts cognitive investment when social complexity is defined operationally as the statistical model composed of social variables that can best predict cognitive investment. On the other hand, if social variables cannot predict cognitive investment, this can also only be taken as tentative evidence against the SBH because it may be that the correct social variables were not included in the analysis. Thus, exploratory analysis may help identify social variables that are most likely to contribute to social complexity, but it cannot be used to distinguish between competing hypotheses when there are no well-defined *a priori* hypotheses. A researcher could instead ask whether their *a priori* social metrics are better predictors of cognitive investment than their *a priori* ecological metrics within a statistical framework, but in this case, the specification of the statistical model can have a large impact on the outcome of the comparison (Hooper et al., 2022). Once again, the underlying issue is that without some independent way to quantify social complexity, it is difficult to statistically infer whether social complexity drove cognitive evolution in the evolutionary past.

Although exploratory analyses are unsuitable for testing the SBH directly, they can still help identify variables of interest and may help us understand social complexity on a more mechanistic level. Social variables, or combinations of social variables, that are strong predictors of cognitive investment can be selected as candidate variables that warrant additional study. The easiest systems to explore will be ones in which species have very similar ecologies and a relatively low-dimensional social environment. Analysis will also be easier if the relevant socioecological variables in the taxon of interest are not highly correlated, allowing statistical methods to disentangle which variables are the best predictors.

*Building experimental support for proxies of social complexity*

Since creating a *de novo* quantitative measure of social complexity is so difficult, experimental approaches to studying social complexity may help infer functional measures of social complexity. To provide independent support for a metric of social complexity, we would need direct evidence that suggests that our metric of social complexity represents the underlying cognitive challenge created by the social environment. Following our thought experiment, we could imagine a hypothetical experiment in which the proportion of times an individual engaged in hawk-hawk contests for resources is measured across individuals reared in different social environments. In this hypothetical experiment, we would be trying to identify the aspects of the social environment that make it harder for individuals to properly assess conspecifics to better understand how the social environment affects cognitive performance. For example, if the interaction partners of an individual were experimentally manipulated, we could test whether the probability of an incorrect hawk-dove decision increased for individuals with more interaction partners. This would help indicate whether an increase in interaction partners could drive demand for more memory. While real empirical research of this nature is no doubt challenging and fraught with its own set of obstacles, it may nonetheless represent a promising future direction of SBH research.

Experimental work focused on identifying relevant cognitive tasks and how performance on those tasks changes as a function of the social environment can help us infer which aspects of the social environment generate social complexity in different species. In this style of experiment, individuals in controlled social environments can be tested on the same cognitive task to establish whether performance declines with some aspects of the social environment. To extend our thought experiment to a more realistic scenario, we can imagine that a species uses some alternative cognitive mechanism to determine social hierarchies instead of our memory mechanism. If we imagine a species that determines hierarchical positions by visually inspecting the size of conspecifics, then we may expect that hierarchies will be less stable when individuals become harder to distinguish visually (*i.e.* equal size; Ang & Manica, 2010a; Ang & Manica, 2010b), but that the number of different individuals an individual interacts with will not have a direct effect on hierarchy formation *per se*. However, for a species that establishes or maintains hierarchies by remembering its position with respect to the identity of other individuals (Tibbets, 2002), we would expect that an individual exposed to many different individuals would be more likely to misremember its position in the hierarchy relative to others. Advancements in our understanding of the cognitive mechanisms that underly a particular behavior are particularly important because some species may be able to use simple cognitive mechanisms to cope with socially complex environments (Barret, et al. 2007).

Unfortunately, finding taxa suitable for this style of experiment is difficult for logistic and ethical reasons, especially since the results would only be straightforward when the species share similar cognitive and behavioral mechanisms. Nevertheless, we believe more progress in cognitive research focused specifically on how the social environment affects performance on cognitive tasks will help us ultimately infer how the social environment creates social complexity. Using empirically determined proxies of social complexity may help us make stronger *a priori* tests of the SBH and complement more traditional comparative research, while also allowing us to move towards more mechanistic understandings of the SBH.

**SUMMARY AND RECOMMENDATIONS**

Although social complexity has garnered significant research interest, a consistent definition of social complexity has been elusive because researchers have traditionally used their own judgement when deciding whether a social environment or behavior was indicative of social complexity. We introduced a thought experiment to provide a tangible example of a system for which it is easier to intuit what social complexity must mean when testing the social brain hypothesis. Our conceptual definition focuses on how aspects of the social environment can create selection for increased cognitive investment by decreasing performance on a cognitive task.

Using this conceptual framework for social complexity may help clarify the conditions under which a metric of social complexity is suitable for testing the SBH, but it also highlights how difficult it is to actually quantify social complexity. Even in our simple thought experiment, we were unable to derive a mathematical expression for social complexity. When analyzing simulated data generated from modeling our thought experiment, we found that univariate proxies of social complexity were vulnerable to misinterpretation because they ignored interactions, non-linear relationships, and possible multicollinearity. Similarly, we were unable to make an *a priori* multivariate measure of social complexity because we could not determine how different variables should be combined to scale with the cognitive challenge imposed by the social environment. This underscores that while there are many different approaches and operational definitions that can be used to create proxies of social complexity, it is unclear if any of these proxies sufficiently capture social complexity as defined in an evolutionary context. Inappropriate proxies of social complexity will tend to underestimate the correlation between social complexity and cognitive investment in cases where the SBH is true. This may partially explain why the results of SBH analyses are sometimes inconsistent and sensitive to how alternative hypotheses are formulated. Thus, we recommend that researchers are careful in their presentation of their analyses and provide statistical and evolutionary context for their measure of social complexity.

In spite of our inability to quantify social complexity in our thought experiment, exploratory statistical analyses and manipulative experiments may improve our mechanistic understanding of social complexity. Empirical research that is conducted in parallel to comparative analyses may allow SBH research to expand into more direct studies of cognition. Simple systems characterized by easily manipulated social environments, species that occupy a range of social environments, and low variability in ecological parameters between species will be particularly useful for more mechanistic SBH research. When researchers can more confidently rationalize or validate their measures of social complexity, the SBH will be more amenable to statistical tests and it will be easier to properly distinguish between alternative hypotheses. Ultimately, recognizing the complexity of social complexity may be a necessary step in the progression of SBH research.

# Supplemental Information

To generate hypothetical data, we simulated 200 unique species, where a species was fully defined by its combination of network metrics. Our species were made by creating all possible combinations of numbers of individuals ($N$), interaction partners ($P$), and interactions per individual ($I$), where $N$ ranged from 10 to 100 in increments of 10, $P$ ranged from four to 64 in powers of two, and $I$ ranged from 100 to 500 in increments of 100. We assumed that resources had a value of one, winners of hawk-hawk interactions paid a cost of 0.5 for the conflict, and that losers of hawk-hawk interactions paid a cost of 1.5 for the conflict. At the start of each simulation, each individual was initialized with a memory, $m$, taken from a Poisson distribution ($\Lambda = 1$) to ensure that memories were non-negative. All $N$ individuals were spatially arranged in a circle and each individual was connected to the $P/2$ nearest individuals to their left and right. We then simulated a generation by playing ($N \times I$)/2 hawk-dove games in the network by randomly choosing ($N \times I$)/2 pairs of partners in the network with replacement. After all hawk-dove games were played, we calculated the fitness, $F$, of an individual as their net payout from their hawk-dove games minus the metabolic cost of their memory. All individuals with a negative fitness were given a fitness of zero and then we calculated the relative fitness for each individual, $RF$, by dividing its fitness by the total fitness of all the individuals in the network.

      To create the next generation of individuals, we chose the memory of $N$ individuals from the current generation with replacement, with the probability of an individual being chosen equal to their $RF$. These $N$ memories were then each mutated with a probability of 0.1, with a mutation event equally likely to increment or decrement the memory by 1 and memory never taking a negative value. Thus, the new generation was created by simulating evolution via selection and mutation. The new individuals were assigned a random $RHP$ and the process repeated. By simulating 15,000 generations of memory inheritance and mutation, we expected that the final $m$ distribution would be approximately equal to the evolutionarily stable strategy (ESS) value of $m$ for a given $N$, $I$, and $P$. To

estimate the ESS memory of a species, we took the average memory of all individuals in the population for the last 500 generations, creating a single ESS memory for each population. Post-hoc perturbation simulations confirmed that 15,000 generations were sufficient for populations to approach their ESS memory.

For the following analysis, we set the metabolic cost of memory, *c*, to 0.1. Though prior analyses confirm that as *c* increases, the ESS memory of a species decreases, supporting the fact that brain evolution is only possible if the costs are sufficiently low relative to available resources (Fig. 1D, Aiello & Wheeler, 1995). To mimic a test of the SBH using univariate proxies of social complexity, we performed linear regressions between ESS memory and each social network parameter. We then repeated those analyses when some of the data were intentionally omitted to show the sensitivity of the results when the species no longer spanned the full range of values for each dimension of the social environment. All analysis and simulations were conducted using Python (Version 3.11.4) (Van Rossum et al., 1995), the scipy.stats package (Virtanen et al., 2020), and the statsmodels package (Seabold & Perktold, 2010).